\documentclass[table,xcdraw,review,preprint]{elsarticle}
\usepackage{cite}
\usepackage{graphicx}
\usepackage{framed}
\usepackage{amsfonts}
\usepackage{subfigure}
\PassOptionsToPackage{subfigure}{tocloft}
\usepackage{fixmetodonotes}
\usepackage{wrapfig}
\usepackage{tabu}
\usepackage[table,xcdraw]{xcolor}
\usepackage{array}
\usepackage[dvipsnames]{}
\usepackage[table,xcdraw]{}
\usepackage{wasysym}
\usepackage{booktabs}
\newenvironment{thisnote}{\par\color{black}}{\par}

\usepackage{graphicx}

\usepackage{hyperref}
\usepackage[modulo]{lineno}
\usepackage{ gensymb }
\usepackage{ tipa }

\usepackage{tikz}
\newcommand{\ballnumber}[1]{\tikz[baseline=(myanchor.base)] \node[circle,fill=.,inner sep=1pt] (myanchor) {\color{-.}\bfseries\footnotesize #1};}

\usepackage{amssymb}
\usepackage{balance}
\usepackage{tipa}
\usepackage{verbatim}
\usepackage{multirow}
\usepackage{array, booktabs, multirow}
\graphicspath{ {images/} }

\usepackage{cellspace}
\usepackage{enumitem}

\usepackage{authorbiography}%
\begin{document}

\begin{frontmatter}
\title{Communication-efficient Certificate Revocation Management for Advanced Metering Infrastructure and IoT}

\author{Mumin Cebe\fnref{myfootnote1}}
\author{Kemal Akkaya\fnref{myfootnote}}
\fntext[myfootnote1]{Computer Science Department, Marquette University, Milwaukee, WI}
\fntext[myfootnote]{Department of Electrical and Computer Engineering, Florida International University, Miami, FL}

\begin{abstract}
\begin{thisnote}
Advanced Metering Infrastructure forms a communication network for the collection of power data from smart meters in Smart Grid. As the communication between smart meters could be secured utilizing public-key cryptography, however, public-key cryptography still has certain challenges in terms of certificate revocation and management particularly related distribution and storage overhead of revoked certificates. To address this challenge, in this paper, we propose a novel revocation management approach by utilizing cryptographic accumulators which reduces the space requirements for revocation information significantly and thus enables efficient distribution of such information to all smart meters. We implemented the proposed approach on both ns-3 network simulator and a testbed. We demonstrated its superior performance with respect to traditional methods for revocation management.
\end{thisnote}
\end{abstract}

\begin {keyword}
Smart Grid Security; Critical Infrastructures Security; Cyber-Physical Systems; Certificate Revocation; Cryptographic Accumulators; Authentication-Authorization-Accounting and Key management  
\end{keyword}
\end{frontmatter}

\section{Introduction}

\begin{table}[t]
\centering
\resizebox{\textwidth}{!}{
\begin{tabular}{llll}
\multicolumn{4}{c}{\textbf{List of Abbreviations}}                                                               \\ \hline
\rowcolor[HTML]{EFEFEF} 
\textbf{AMI} & Advanced Metering Infrastructure & \textbf{HMS}  & Head-End Management Server                     \\
\textbf{CA}  & Certificate Authority            & \textbf{MAC}  & Medium Access Control                          \\
\rowcolor[HTML]{EFEFEF} 
\textbf{CRL} & Certificate Revocation List      & \textbf{OCSP} & Online Certificate Status Protocol             \\
\textbf{DHT} & Distributed Hash Table           & \textbf{NIST} & National Institute of Standards and Technology \\
\rowcolor[HTML]{EFEFEF} 
\textbf{DMZ} & Demilitarized Zone               & \textbf{PKI}  & Public Key Infrastructure                      \\
\textbf{HES} & Head-End System                  & \textbf{RSA}  & Rivest–Shamir–Adleman 
\\ \hline
\end{tabular}}
\end{table}

The existing power grid is currently going through a major transformation to enhance its reliability, resiliency, and efficiency by enabling networks of intelligent electronic devices, distributed generators, and dispersed loads \citep{farhangi2010path}, which is referred to as \textit{Smart(er) Grid}. Advanced Metering Infrastructure (AMI) network is one of the renewed components of Smart Grid that helps to collect smart meter data using a two-way communication\citep{saputro2012}. Smart meters and integrated Internet-of-Things (IoT) devices are typically connected via a wireless mesh network with a gateway (or access point) serving as a relay between the meters and the utility company.

The security requirements for the AMI network are not different from the conventional networks as confidentiality, authentication, message integrity,  access control, and non-repudiation are all needed to secure the AMI. Confidentiality is required to prevent exposure of customer's private data to unauthorized parties while integrity is necessary to ensure that power readings are not changed for billing fraud. Furthermore, authentication is crucial to prevent any compromised smart meters communicating with other smart meters. On the other hand, the National Institute of Standards and Technology (NIST) urges to use Public Key Infrastructure (PKI) for providing the security requirements of AMI \citep{nist2014}. As an example, companies such as Landis\&Gyr and Silver Spring Networks already use PKI to provide security for millions of smart meters in the US\citep{landisgyr}. In such a PKI, the public-keys for smart meters and utilities are stored in \textit{certificates} which are issued by Certificate Authorities (CAs). The employment of PKI in AMI requires management of certificates which include the creation, renewal, distribution and revocation. In particular, the certificate revocation and its association with smart meters are critical.
\vspace{10pt}
\begin{thisnote}
\subsection{Problem Description and Existing Solutions}
Several reasons necessitate revoking certificates, such as key compromise, certificate compromise, excluding malicious meters, renewing devices, etc.  Besides, if there is a vulnerability in the algorithms or libraries that are used in certificate generation, a massive number of revocations may additionally occur. For instance, a recent discovery of a chip deficiency on RSA key generation caused revocation of more than 700K certificates of devices that deployed this specific chip \citep{2017ccsnemec} and renowned heartbleed vulnerability caused the revocation of millions of certificates, immediately \citep{durumeric2014matter}.  Thus, to establish secure communication, a smart meter should check the status of the other smart meter's certificate against a certificate revocation list (CRL) that keeps all revoked certificates. Considering the large number of smart meters in an AMI and the fact that the expiration period can be even lifelong in particular applications \citep{landisgyr}, the CRL size will be huge. Consequently, revocation management becomes a burden for the AMI infrastructure which is typically restricted in terms of bandwidth. This overhead is particularly critical since the reliability and efficiency of AMI data communication are crucial for the functionality of the Smart Grid. Considering the potential impact on the performance of AMI applications \citep{mahmoud2015investigating}, handling the overhead of revocation management is essential.

Certificate revocation management is commonly handled by utilizing CRL that is stored in the smart meter. The status of a smart meter is determined by checking whether its certificate is listed in the CRL or not. An alternative method would be to store the CRL in a remote server as in the case of Online certificate status protocols (OCSPs) \citep{galperin2013x}\citep{pettersen2013transport}. In OCSP, an online and interactive certificate status server stores revocation information. Thus, each time a query is sent to the server to check the status of the certificate. While OCSP-like approaches can be advantageous on Internet communications, employing them for AMI is not attractive since it will require access to a remote server for each time. In this regard, another alternative would be to use OCSP \textit{stapling} \citep{pettersen2013transport} where the smart meters query the OCSP server at certain intervals and obtain a signed timestamped OCSP response which 
is included ("stapled") in the certificate.  Again, this approach also needs frequent access to a remote server. Moreover,  the 'stapled' certificates should be downloaded frequently by smart meters to ensure security, and this will create additional traffic overhead on the AMI which affects applications such as demand response or outage management. 

\subsection{Our Approach and Contributions}
In this paper, we propose a communication-efficient revocation or CRL mananegment scheme for AMI networks by using RSA accumulators\citep{camenisch2002dynamic}. RSA accumulator is a cryptographic tool which is able to represent a set of values with a single accumulator value (i.e., digest a set into a single value). Also, it provides a mechanism to check whether an element  is in the set or not which implicitly means that cryptographic accumulators can be used for efficient membership testing. Due to the attractiveness of size, in this paper, we adapt RSA accumulators for our needs by introducing several novel elements as following:

\begin{itemize}
\item An accumulator manager is introduced within the utility company (UC) that is tasked with collection of CRLs from CAs and accumulating these CRLs (i.e., revoked certificates' serial numbers) to a single accumulator value which will then be distributed to the smart meters. 
\item We also introduce a non-revoked proof tuple for allowing a smart meter to check whether another meter's certificate is revoked without referring to the CRL file.
\item We defined additional entities within AMI and assign functions to them to govern an accumulator based revocation management.
\item We introduced several security countermeasures against possible attacks to a accumulator-based scheme.
\end{itemize}

The computation and communication related aspects of the proposed approach is assessed via simulations in ns3 network. In addition, we built an actual testbed using in-house smart meters to assess the performance realistically. We compared our approach with the other methods that use conventional CRL schemes and Bloom-filters  \citep{akkaya2014efficient}. The results show that the proposed approach significantly outperforms the other existing methods in terms of reducing the communication overhead that is measured with the completion time. The overhead in terms of computation is not major and can be handled in advance within the utility that will not impact the smart meters.

This paper is organized as follows: In the next two sections, we summarize the related work and the background. Section IV introduces the threat model. Section V presents the proposed approach with its features. Section VI and VII are dedicated to evaluation criteria and experimental validation. Section VIII analyzes the security of the approach. Section IX discusses the benefits and limitations. The paper is concluded in Section X. 

\end{thisnote}

\label{intro}


\section{Related Work} \label{section2}

\begin{thisnote}
Due to increasing threats towards Smart Grid, there has been a number of efforts to adapt PKI for Smart Grid communication infrastructure. For instance, Metke et al.\citep{metke2010security} surveyed the existing key security technologies in Smart Grid domain by mainly focusing on PKI. On the other hand, the study  \citep{mahmoud2013efficient} stressed the importance of revocation overhead of PKI in Smart Grid. Beyond directly related studies on the PKI and Smart Grid relation, we also focus on studies about cryptographic accumulators and membership management. In this section, we examine the relation between this study and previous studies and highlight major differences.

\subsection{Revocation Management in AMIs}
The studies \citep{mahmoud2015investigating} investigated different revocation management aspects such as short-lived-certificate scheme, tamper-proof device scheme, Online Certificate Status Protocol (OCSP), conventional CRL, and compressed CRL. However, this study just hypothetically analyzed the applicability of existent revocation solutions for AMI.  The first offered approach that focused on reducing the revocation management overhead for AMI was based on Bloom Filters \citep{rabieh2015scalable}. They provided a Bloom Filters based scheme particularly to reduce the size CRL. \end{thisnote} However, since Bloom Filters suffer from false positives, the approach
requires accessing the CA to check the validity of a certificate. Our proposed scheme, on the other hand, never
requires accessing a remote server and provides a better reduction
on CRL size. The study in \citep{cebe2017efficient} use distributed hash tables (DHT) to reduce the CRL size again. Although this study provides a reduction in CRL size, it suffers from additional inter-meter communication overhead for accessing the CRL information. 
\begin{thisnote}
We would like to note that a very preliminary version of this work was published in \citep{cebe2018efficient}. In this work, we improved the various aspects of the previous one. First, we improved computation performance utilizing Euler’s Theorem. Second, we extended our threat
model to new attack types that were not considered in the conference version. In this regard, we changed our approach in several ways: We proposed to use an initial secret during accumulation. We then introduced a non-revoked proof concept that was not used before in any of the revocation works. This required major changes to the accumulation process which was not in \citep{cebe2018efficient}. We finally proposed an extensive certificate verification protocol as countermeasures to the new threats. This also required proposing a new secure multi-level AMI architecture as opposed to the monolithic architecture used in \citep{cebe2018efficient}. In addition, we added several new experiments with accumulator computation overhead under various assumptions.
\end{thisnote}
\subsection{Cryptographic Accumulators}
Benalog and DeMare \citep{benaloh1993one} first introduced cryptographic accumulators. After their first appearance, there have been studies \citep{camenisch2002dynamic,reyzin2016efficient,baldimtsi2017accumulators} offering to use them for membership testing. However, these studies solely focused on building the cryptographic fundamentals of accumulators, and thus, omit application-specific issues and security features when deploying them. Besides, these studies are offering to use accumulators for membership testing by accumulating a valid list. Considering AMI, accumulation of valid smart meter's certificates to provide a revocation mechanism would constitute a significant overhead due to the fact that revocation frequency is less than that of creating new certificates {(i.e., no need to update the accumulator each time when a new smart meter is added to AMI)}. Furthermore, since the number of revoked certificates is also less than the number of valid certificates which affects the required computation time significantly\citep{durumeric2014matter}. 
Our approach mitigates these drawbacks by addressing security and application-specific issues and offering to use CRLs instead of valid certificates.


\section{Preliminaries}
\label{section3}
\begin{thisnote}
Before explaining  our  approach  we  provide  some  cryptographic background of accumulators and its particular form as RSA accumulators. In addition, to help the reader grasp a general idea of revocation management through CRLs, we explain the CRL and delta-CRL notions.
\end{thisnote}
\subsection{Background on Cryptographic Accumulators}
Benaloh and De Mare\citep{benaloh1993one} introduced the cryptographic accumulator concept which is a one-way hash function with a special property of being \emph{quasi-commutative}. A quasi-commutative function is a special function $\mathcal{F}$ such that $y_0,y_1,y_2 \in \mathbb{Y}:$
\begin{equation}
\mathcal{F}(\mathcal{F}(y_0,y_1),y_2)=\mathcal{F}(\mathcal{F}(y_0,y_2),y_1)  \label{quasi}
\end{equation}
The properties of this function can be summarized as follows: \textit{1)} it is a one-way function, i.e., hard to invert; \textit{2)} it is a hash function for obtaining a secure digest $\mathcal{A}$ (i.e., accumulator value) where $\mathcal{A} = \mathcal{F}(\mathcal{F}(\mathcal{F}(y_0,y_1),y_2),...,y_n)$ for a set of values $\{y_0,y_1,y_2, . . . , y_n\} \in \mathbb{Y}$; \textit{3)} it is a \emph{quasi-commutative} hash function which is different from other well-known hash functions such that the accumulator value $\mathcal{A}$ does not depend on the order of $y_i$ accumulations. 

These properties allow cryptographic accumulators to be used for a condensed representation of a set of elements. In addition, since the resulting accumulated hashes of $y_i$ ($\mathbb{Y}=\{y_i;~0<i<n\}$) 
stays the same even if the order of hashing is changed, it can be used for efficient membership testing by using a special value called witness value $w_i$. For instance, the witness $w_i$ of corresponding $y_i$ is calculated by accumulating all $y_j$ except the case where $i \neq j$ (e.g., $w_i = \mathcal{F}(\mathcal{F}(\mathcal{F}(y_0,y_1),...,y_{j-1},y_{j+1}...,y_n)$). Then, when necessary any of the members can check whether $y_i$ is also a member of the group by just verifying whether $\mathcal{F}(w_i,y_i)=\mathcal{A}$. Note that, because $\mathcal{F}$ is a one-way function, it would be computationally infeasible to obtain $w_i$ from $y_i$ and $\mathcal{A}$. 
However, there is a risk for collusion in this scheme when an adversary can come up with ${w_i}^{'}$ and ${y_i}^{'}$ pairs where ${y_i}^{'} \notin \mathbb{Y}$ to obtain the same accumulator value: $\mathcal{F}({w_i}^{'},{y_i}^{'})=\mathcal{A}$. In the literature, there is already a cyrptographic accumulator, namely the RSA construction \citep{baric1997collision} which guarantees that finding such pairs is computationally hard by restricting the inputs to the accumulator function to be prime numbers only. This scheme is known as collision-free accumulator that enables secure membership testing (i.e., without any collision). Therefore, in this paper, we chose to employ RSA construction which is elaborated next. 
\subsection{RSA Accumulator}
RSA accumulator\citep{baric1997collision} has a RSA modulus $\mathcal{N}=pq $, where $p$ and $q$ are strong primes. The RSA accumulation value $\mathcal{A}$ is calculated on consecutive modular exponentiation of prime numbers set $\mathbb{Y} = \{y_1,...,y_n\}$ and $g$ is quadratic residue of $\mathcal{N}$ as follows:
\begin{equation}
\mathcal{A} = g^{y_1,...,y_n}~(mod~\mathcal{N})   \label{eq:accumulator}
\end{equation}
The witness $w_i$ of corresponding $y_i$ is calculated by accumulating all values 
except $y_i$:
\begin{equation}
w_i = g^{y_1,...,y_{i-1},y_{i+1},...,y_n}~(mod~\mathcal{N})   \label{eq:witness}
\end{equation}
Then, the membership testing can be done via a simple exponential operation by comparing the result with the accumulator value $\mathcal{A}$:
\begin{equation}
    w_i^{y_i} \leftrightarrow \mathcal{A}
    \label{eq:authenticate}
\end{equation}
The described accumulator scheme so far basically allows generation
of a ``witnesses" to prove that an item is in the set. A more advanced accumulator would
offer proofs of non-membership which proves that an item is \textbf{NOT} in the set \citep{li2007universal}. For this scheme, let us assume any $x \notin \mathbb{Y} = \{y_1,...,y_n\}$. In a nutshell, the non-witness values can be computed by the following steps: Let $u$ denote $\prod_{i=1}^ny_i$, the scheme finds  non-witness $nw_1,b$ value pairs of $x$ by solving the equation of $nw_1\times u+b \times x = 1$ using the Extended Euclidean algorithm. Then, the scheme computes an additional value $nw_2$ such that:
\begin{equation}
{nw_2}=g^{-b}~(mod~\mathcal{N})
\label{eq:witnessCalculation}
\end{equation}
After these steps, the item $x$ will have cryptographic proof values $nw_1$ \& $nw_2$ which can be used to ensure that the item $x$ is \textbf{NOT} in the set $\mathbb{Y}$. Then, any third party that posses the $\mathcal{A}$ value can do the non-membership test of $x$ via a simple exponential operation by checking whether the following equation holds:
\begin{equation}
\mathcal{A}^{nw_1}  \leftrightarrow {nw_2}^x \times g~(mod~\mathcal{N})
 \label{eq:authenticate2}
\end{equation}
Besides, if a new value $y^{'}$ is added to list, the accumulator value is updated by using the previous accumulator value $\mathcal{A}$:
\begin{equation}
\mathcal{A}^{'}  = 
\mathcal{A}^{y^{'}}~(mod~\mathcal{N})
 \label{eq:updateAccumulator}
\end{equation}

\subsection{Certificate, CRL and Delta CRLs}

As we deal with certificates, we would like to also provide some basic background on certificates and their management. Certificates are issued by a CA with a planned lifetime to an expiration date and have unique serial numbers. 
Once issued, these certificates are valid until their expiration date. However, there are various reasons that cause a certificate to be revoked before the expiration date. These reasons include but not limited to compromise of the corresponding private key, changing the underlying device infrastructure, etc. 

Revocation causes each CA regularly issued a signed list called a CRL which is a time-stamped list consisting of serial numbers of revoked certificates and revocation dates. When a PKI-enabled system uses a certificate (for example, for verifying the integrity of a message), that system should not only check the time validity of the certificate, but an additional check is required to determine a certificate's revocation status during the integrity check. To do so, CRL can be checked to determine the status of the certificate.

There are two main types of CRL: \textit{full CRLs} and \textit{delta CRLs}. A full CRL contains the status of all revoked certificates which are not expired yet. \textit{Delta CRLs}, which is a concept defined in in RFC 5280  \citep{cooper2008internet}, contain only the status of newly revoked certificates that have been revoked after the issuance of the last \textit{full CRL} and before the new release of it. Therefore, a \textit{full CRL} is issued for a limited time frame and should be updated regularly. Until next update time, \textit{delta CRLs} help keeping track of the newly revoked certificates. When \textit{delta CRLs} are enabled, the CA can distribute \textit{full CRLs} at longer intervals (for reducing distribution overhead) and delta CRLs at shorter intervals. An important point about delta CRL concept is that it does not eliminate the requirement of full CRL distribution. The full CRL must still be re-distributed when the previous full CRL expires since CRL has also a lifetime period as certificates and the lifetime period of delta CRLs are dependent on the lifetime of the previous full CRL. This means both the \textit{full CRL} and \textit{delta CRL} should be updated regularly by all the potential nodes that will be using them. In the case of AMI, these CRLs may contain thousand of revoked certificate IDs due to longer expiration dates of issued certificates (even lifelong \citep{landisgyr}) and need to be distributed to the each smart meters which will cause a huge overhead due to their size as will be shown in the experiments.





\section{System and Threat Model}
\label{sec:threat}

\begin{thisnote}

\begin{figure}[!ht]
\begin{centering}
\fbox{
\includegraphics[scale=0.4]{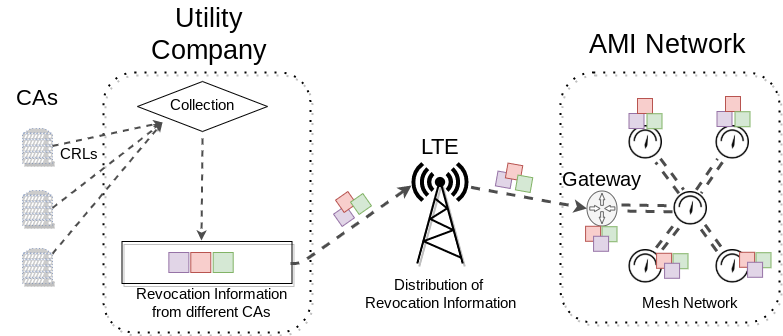}}
\caption{System Model}
\vspace{-0mm}
\label{fig:initialsystem}
\end{centering}
\end{figure}

In this paper, we build a revocation management scheme for a typical AMI infrastructure. Basically, revocation information is collected by the utility company in the forms of CRL files. Each CRL file contains revoked certificates IDs issued by different CAs. Then, all these revocation information are disseminated to AMI through a 4G/LTE and AMI mesh communication infrastructure. A sample system model is shown in Fig. \ref{fig:initialsystem}. 

The security of the proposed revocation management scheme depends on the secure implementation of the proposed accumulator-based system. Therefore, we consider the following threats to the security of the proposed approach and identified the relevant security goals. Note that in our attack model, we assume that both the accumulation process within the perimeter and smart meters (outside the perimeter) can be compromised. Besides,  the communication between UC and smart meters is happening on a non-secure medium which means an adversary can eavesdrop the communication both actively and passively. This threat model is very strong and adequate to represent the increasing threats to Smart Grid.
However, a single counter-measure against this threat model would not be sufficient when considering the broad and diverse attack surface of it. To ensure the security of AMI against adversaries, the utility company needs to deploy intrusion prevention systems and proper attack prevention tools as well. Thus, we assume that a PKI inspection system along with an intrusion detection system (IDS) is already deployed and provides device-level controls to protect PKI keys and informs UC in case of any infiltration.
\end{thisnote}

\begin{itemize}[label=\null]
\begin{thisnote}
\item \ballnumber{1} \textbf{Compromising Smart Meters:} In an attacker's perspective, the meter/gateway is the entry point to the AMI. The attacker can use a compromised smart meter or impersonate the gateway to apply various attacks.
\item \ballnumber{2} \textbf{Compromising the UC Servers:} Apparently, compromising the servers within perimeters of UC provides lots of attack opportunities to adversary. The adversary can target AMI by directly attacking revocation management through compromising servers that governs revocation operations.
\item \ballnumber{3} \textbf{Compromising the Communication:} When UCs are deploying AMI systems, they generally opt-out enabling encryption since IEEE standards does not enforce the UCs to deploy encryption due to various reasons \citep{IEEE1815}. It makes AMI open to adversaries who can easily eavesdrop whole AMI traffic or a portion of it. This can also pose a threat to revocation management.
\end{thisnote}
\end{itemize}

\section{Proposed Approach}
\label{sec:proposed}
\subsection{Overview}
The proposed approach basically eliminates the need to store and distribute CRLs when the devices communicate in a secure manner. Instead of keeping a CRL file for verification of revocation status of certificates, our approach dictates to store at each device (e.g., smart meter, gateway, HES, etc.)  only an accumulator value and a proof which proves the validity of the device's certificate. The accumulator value and proof can be computed at the utility company and distributed to devices in advance. Any updates regarding revoked certificates trigger re-computation of these values. Keeping just two integer values for revocation management brings a lot of efficiency in terms of storage and distribution overhead as will be shown in the Experiments section. In the next subsections, we will explain the details of our approach.
\subsection{Adaptation of RSA Accumulator for Our Case}
\begin{thisnote}
To apply the cryptographic accumulators for revocation management, the revocation management needs to be viewed holistically from the lens of systems thinking to ensure security. We took a bottom-up approach while adapting the accumulator scheme to our approach. First, we modified the CRL inputs to meet the requirements for constructing a secure accumulator setup.  Second, we improved the performance of accumulator calculation. Third, the accumulation process was divided into different functions and  their tasks were defined. Then, we introduced new entities to AMI and assigned tasks to them. Lastly, we constructed a revocation check protocol that utilizes the produced accumulator solution. This section covers how we accomplished all these steps in details.
\end{thisnote}
\begin{itemize}

\item[\textbf{a.}] {\textbf{Integration of CRL and non-witness Concept:}}
    
In the traditional CRL approach, when a smart meter presents its certificate to the recipient meter, that meter needs to verify that the presented certificate is \textbf{NOT} in the CRL. 
To be able to employ the accumulator approach, we generate \textit{non-witness values} for the presenter to prove that it is not in the list. 
We accumulate the revocation information (stored in CRLs) into a single accumulator value and produce non-membership witnesses for the non-revoked smart meters.



\item[\textbf{b.}] {\textbf{Reducing the Complexity of Accumulator Computation:}}

While computing the accumulator value using Eq.~\ref{eq:accumulator}, the exponent needs to be computed as
$\prod_{i=1}^ny_i$ before doing the modular exponentiation. This becomes infeasible when the size of $\mathbb{Y}$ increases since $u=\prod_{i=1}^ny_i$ will be $n \times k$ bits assuming each $y_i$ is a $k$-bit integer. In our approach, we decided to use Euler's Theorem \citep{rosen2005elementary} to cope with this complexity. With access to the totient of $\mathcal{N}$ (i.e., $\phi(\mathcal{N})$), the exponent of $g$ in accumulation computation will be $u^{'}=\prod_{i=1}^ny_i~mod~ \phi(\mathcal{N}$). Thus, with the knowledge of the totient, it becomes more efficient to compute the required values via reducing the $u$ by $\phi(\mathcal{N})$. 

\item[\textbf{c.}] {\textbf{Generating Prime Inputs for the Accumulator:}}

For accumulation, we can use the certificate IDs ($c_{id}$) which are generated by the CAs. However, to ensure a collision-free accumulator, we need to use only prime numbers as dictated by the RSA accumulator. Since CRLs contain arbitrary serial numbers for certificate IDs, it is necessary to compute a prime representative for each certificate ID as an input to the RSA accumulator. Thus, we used the random oracle based prime number generator described in \citep{papamanthou2008authenticated} to obtain prime representatives of certificates from their serial numbers ($c_{id}$). The scheme basically has a random oracle $\ohm()$ function which produces a random number $r$ for an input $c_{id}$. We use $\ohm()$ to find a \textit{256-bit} number, $d$, which causes the result of the following equation to be a prime number:
\begin{equation}
    y = 2^{256}\times\ohm(c_{id})+d
\end{equation}
By solving this equation, we generate a prime representative $y$ for a revoked certificate. The reader is referred to \citep{papamanthou2008authenticated} for security proof details of the method.


\item[\textbf{d.}] {\textbf{Defining Functions of Revocation Management:}} \\
After preparing the inputs, we compiled and modified the offered accumulator structure and proposed the following functions to construct revocation management for AMI. Our RSA accumulator uses the following input sets: $\mathbb{Y}$ \textit{is the set of prime representatives of revoked certificates' serial numbers} and  $\mathbb{X}$ \textit{is set of prime representative of valid certificates' serial numbers} where $x \in \mathbb{X}$ :

\begin{itemize}
\item $aux_{info}, \mathcal{N} \leftarrow Setup(k)$: This function is to setup the parameters of the accumulator. It takes $k$ as an input which represents the length of the RSA modulus in bits (e.g., 2048, 4096, etc.) and generates modulus $\mathcal{N}$ along with $aux_{info}$ which is basically Euler's totient $\phi(\mathcal{N})$. 


\item $\mathcal{A} \leftarrow ComputeAcc(\mathbb{Y},r_k,aux_{info})$: This is the actual function which accumulates revocation information by taking prime representatives of serial numbers set $\mathbb{Y}$. 
While computing the accumulator value, we propose to use an initial random secret prime number $r_k$ as a first exponent ($g^{r_k}$) in Eq.~\ref{eq:accumulator}. 
\item $nr_{proof} \leftarrow ComputeNonRevokedProof(aux_{info},\mathbb{Y},x)$: This function first computes a pair of non-witness values represented as $(nw_1,nw_2)$ for a valid certificate whose prime representative is $x$. Then, the UC concatenates the non-witness value pair with $x$ and the serial number of the certificate creating a 4-tuple called $nr_{proof}$. 

\item ${0,1} \leftarrow RevocationCheck(\mathcal{A},nr_{proof})$: When a smart meter which has a prime representative $x$ wants to authenticate itself to another party, the other one uses $nr_{proof}$ and $\mathcal{A}$ to verify that $x$ is \emph{not} in the accumulated revocation list by checking Eq.~\ref{eq:authenticate2}.
\item $\mathcal{A}^t \leftarrow UpdateAcc(\mathcal{A}^{t-1},\mathbb{Y}^t)$: This function is for updating the accumulator value $\mathcal{A}$ when the revocation information is updated via deltaCRLs. It  takes a set of prime representatives of corresponding newly revoked certificates $\mathbb{Y}^t$ and latest accumulator value $\mathcal{A}^{t-1}$, and returns the new accumulator value $\mathcal{A}^t$ by utilizing Eq.~\ref{eq:updateAccumulator}. 

\item ${nr_{proof}}^t \leftarrow UpdateNonRevokedProof(\mathcal{A}^{t},\mathbb{Y}^t,x)$: This function is for updating the non-revoked proof of corresponding valid smart meters when the revocation information is updated via deltaCRLs. It takes a set of prime representatives of corresponding newly revoked certificates $\mathbb{Y}^t$, the updated accumulator value $\mathcal{A}^t$, and the prime representative $x$  and returns non-revoked proof ${nr_{proof}}^t$ of smart meter after some additional certificates are revoked by utilizing the process for Eq.~\ref{eq:witnessCalculation}.

\end{itemize} 

\end{itemize}


Next, we define the components of the proposed framework. 
\subsection{Components of Revocation Management System}
We propose the system architecture shown in Figure~\ref{fig:AMI_Inf} to enable the proposed revocation management and to define its interaction with the deployed AMI components. In addition, the newly introduced components of this architecture and their roles in executing the above defined functions are described below: 
\begin{figure}[ht]
   \centering
   \includegraphics[width=0.95\textwidth]{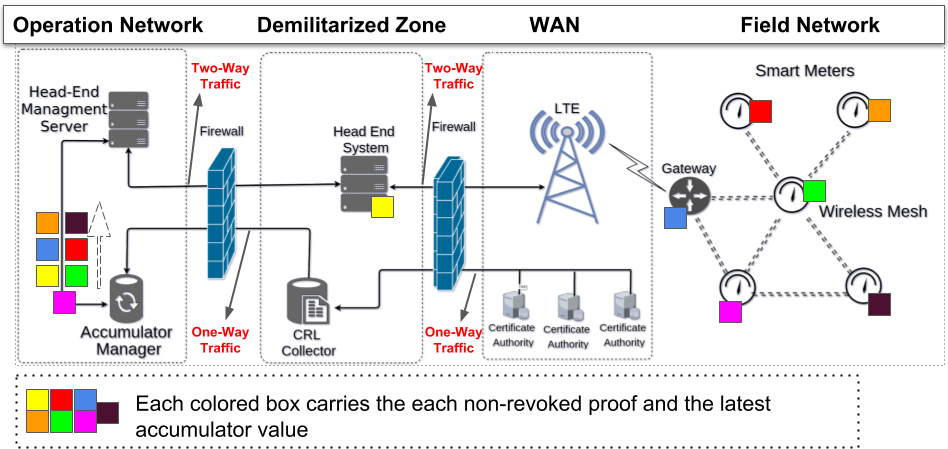}
   \vspace{-10pt}
    \caption{The structure of proposed revocation management.}
    \label{fig:AMI_Inf}
    \vspace{-10pt}
\end{figure}
\begin{itemize}
\item \emph{Smart Meters and Gateway:} The smart meters and gateway can directly communicate with each other and with Head-end System (HES) over LTE. Thus, to ensure the security of applications, these devices need to run the $RevocationCheck()$ function and carry the latest $\mathcal{A}$ and the corresponding $nr_{proof}$.
\item \emph{Head-End System:} HES is an interface between the utility operations center and smart meters, and it is located in  a demilitarized zone (DMZ). The primary function of the HES is collecting the power data from smart meters and transfer them to head-end management servers (HMS). 
Since it has two-way communication with smart meters, it needs to run the $RevocationCheck()$ function and carry the latest $\mathcal{A}$ and its $nr_{proof}$.
\item \emph{CRL Collector:} The CRL collector plays one of the key roles in our revocation management system. It basically collects CRLs from various CAs and feeds them to the Accumulator Manager. Since it has an open interface to the outside network (communicating with other CAs), it is placed in DMZ area. 
\item \emph{Accumulator Manager:} Accumulator Manager is the core of our revocation management scheme. It gets CRL information from the CRL Collector and accumulates them to obtain latest accumulator value. It implements the \emph{Setup()}, \emph{ComputeAcc()}, \emph{ComputeNonRevokedProof()}, \emph{UpdateAcc()}, and \emph{UpdateNonRevokedProof()}  functions. Whenever a new accumulator value is calculated at a time $t$, it sends the accumulator value $\mathcal{A}^t$ and updated ${nr_{proof}}^t$ to the HMS which then forwards them to HES for distributing to the smart meters.


\item \emph{Head End Management Server:} The collected data is managed within HMS. It basically monitors activity logs, identifies new devices and manages incident response processes. As mentioned, the HMS collects the newly generated $\mathcal{A}$ and $nr_{proof}$ values and sends them to HES for distribution. 
\end{itemize}
\subsection{Revocation and Certificate Verification Processes}
In this section, we describe the proposed revocation scheme and the protocol for certificate verification.  

\subsubsection{Accumulating the CRL}
This process includes two phases namely the setup phase and the update phase which are described below.
\begin{itemize}
\item \textbf{The setup phase:} 
In this phase of our approach, the Accumulator Manager in the UC basically accumulates the revoked certificate IDs in \emph{full CRLs}. This process works as follows: The \emph{full CRL} files are read, and each certificate ID and its issuer's public key are concatenated to obtain a unique string that will be input to the accumulator. Note that the issuer's public key is concatenated on purpose to eliminate any duplicates in serial numbers that may come from different CAs. Then, the Accumulator Manager calculates prime representatives for each concatenated string and accumulates these prime representatives to obtain the accumulator value. 
Finally, the Accumulator Manager generates non-revoked proofs (i.e., the 4-tuple $nr_{proof}$) for each end-device (smart meter, gateway, HES, etc.) by using $ComputeNonRevokedProof()$ function.
\item \textbf{The update phase:} This phase is for revocation information updates that can be done through \textit{delta CRLs}. Due to such updates, the accumulator value $\mathcal{A}$ and $nr_{proof}$ values should be updated. To update these values, the Accumulator Manager first prepares the prime representatives for the newly revoked certificates (i.e., the ones that are included in the \textit{delta CRLs}) by following the same approach in the setup phase. It then updates the previously computed accumulator value, $\mathcal{A}^{t-1}$, by using the $UpdateAcc()$ function to obtain $\mathcal{A}^t$ which is then used to generate new $nr_{proof}$ tuples for the end devices by using the $UpdateNonRevokedProof()$ function. 
\end{itemize} 
\begin{figure}[ht]
\vspace{-4pt}
   \centering
   \includegraphics[scale=.35]{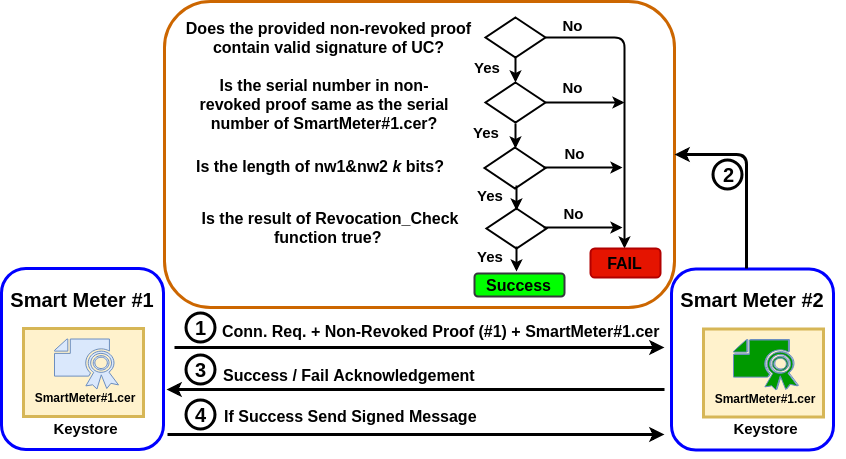}
   \vspace{-10pt}
    \caption{Certificate Verification Protocol Scheme.}
    \label{fig:mutual}
\end{figure}
\subsubsection{Certificate Verification Protocol} 
When two meters communicate by sending/receiving signed messages, the signatures in these messages need to be verified. To be able to start the verification process, a receiving device needs to use the public key (for signature verification) presented in the certificate sent to itself. To ensure that this certificate is not revoked, then it needs to initiate a process which we call as certificate verification protocol.  Figure~\ref{fig:mutual} shows an overview of this process. 
Basically, the receiving device checks the corresponding $nr_{proof}$ tuple's signature to ensure that it is produced by the UC. Once the signature is verified, it then checks whether the the serial number within the tuple is same as the serial number of the provided certificate (i.e., either EndDevice\#1.cer). For additional security, it also checks the length of the $nw_1\&nw_2$ to see whether it is equal to the first accumulation setup parameter $k$. Next, it perfoms $RevocationCheck()$ function amd checks whether the provided $nr_{proof}$ is correct. Finally, the signature of connection request message is checked to ensure the integrity and authenticity of the request. If all these steps are successful, the end-device has successfully complete the certificate verification protocol. Note that, without carrying the $nr_{proof}$, a smart meter can not be authenticated even if it has a valid certificate.



\section{Evaluation of the Approach and its Objectives}
\begin{thisnote}
The main objective of our work is to decrease the dissemination overhead of revocation information on AMI. However,  although any reduction in this overhead is important for the general health of Smart Grid, achieving this goal without sacrificing the security is vital.  Thus, we have determined the following general measures to evaluate the proposed approach in terms of security, communication, computation and storage.
\begin{itemize}
\item First, we will evaluate distributing process of non-revoked proofs to smart meters to assess the communication-related overhead of our approach.
\item Second, since our approach requires computational resources to calculate the non-revoked proofs and accumulator value,  we will evaluate computational costs on UC servers. 
\item Considering the limited computation resource of a typical smart meter, it is essential to evaluate computational aspects on smart meters as well. Moreover, we will evaluate storage space requirement of our approach for smart meters.  
\item Finally, we will assess the security of the proposed approach against threats that are defined in Section~\ref{sec:threat}.
\end{itemize}
We evaluate the communication, computation and storage overhead of our approach by using the following metrics:
\end{thisnote}
\begin{thisnote}

\begin{itemize}
\item \textit{Completion Time}: This metric is defined for communication overhead assessment, which indicates the total elapsed time to complete the distribution of accumulator value and non-revoked proofs to the smart meters from the HES. This metric hints on the communication overhead of revocation management in terms of assessing how it keeps the communication channels busy which are critical for carrying other information.
\item \textit{Computation Time}: This is the metric to measure the total time for completing the required computations such as computation of accumulator value, prime representatives, and revocation check time, etc. 

\item \textit{Storage}: This metrics indicates the amount of space for storing the CRL information in the meters. 
\end{itemize}
For comparison to our approach, we used two other baselines from the literature: 
\begin{itemize}
\item \textit{Traditional CRL Method}: Each smart meter keeps the whole CRL \citep{mahmoud2013efficient} locally which is distributed by the UC. 
\item \textit{Bloom Filter Method}: A Bloom filter \citep{akkaya2014efficient} is used to store revoked certificates information. Note that, we employed \textit{murmur} hash function, which is a non-cryptographic hash function suitable for \textit{fast} hash-based lookup, to build this Bloom filter. In this case, the Bloom Filter is distributed to each meter by the UC.  
\end{itemize}

\end{thisnote}
\section{Performance Evaluation}
\label{sec:result}
\subsection{Experimental Setup}
To assess the performance of the proposed approach, we implemented it in C++ by using FLINT \citep{hart2011flint}, which is the fastest library for number theory and modular arithmetic operations over large integers. 
For the RSA modulus generation and prime representatives computation, we used Crypto++ library since it allows thread-safe operations. 
We prepared a binary-encoded \textit{full CRL} and \textit{delta CRL} that have been digitally signed according to RFC 5280 standard and  
contained 30,000 and 1000 revoked certificates for \textit{full CRL} and \textit{delta CRL} respectively. The \textit{full CRL} was used to compute $\mathcal{A}$ and  $nr_{proof}$ tuples during the setup phase while the \textit{delta CRL} ws used for updating both $\mathcal{A}$ and $nr_{proof}$ tuples.

For communication overhead assessment, we used the well-known ns-3 simulator \citep{ns3} 
which has a built-in implementation of IEEE 802.11s mesh network standard. The underlying MAC protocol used was 802.11g. 
We created two different AMI grid topologies that consist of 81 and 196 smart meters. Even though the number of smart meters in our simulation setup is less than a real AMI setup, it still represents a practical setup in terms of the number of hops due to limited transmission range of 802.11g which leads to multiple hops to reach a smart meter from the gateway (e.g., for 81 nodes the average hop count is 6 and for 196 setup average hop count is 9). In a typical AMI setup in the wild, utilities are able to use 900MHz frequency bands \citep{TinyMesh} which helps to reach thousand of smart meters through a few hops due to the extended transmission range. Unfortunately, ns-3 does not support those frequencies to build a mesh network, and thus we created a simulation environment which reflects similar number of hops as in the wild.    
\begin{figure}[!ht]
\label{fig:smartmeter12}
\begin{minipage}{0.30\columnwidth}
\includegraphics[width=\columnwidth,height=2.2cm]{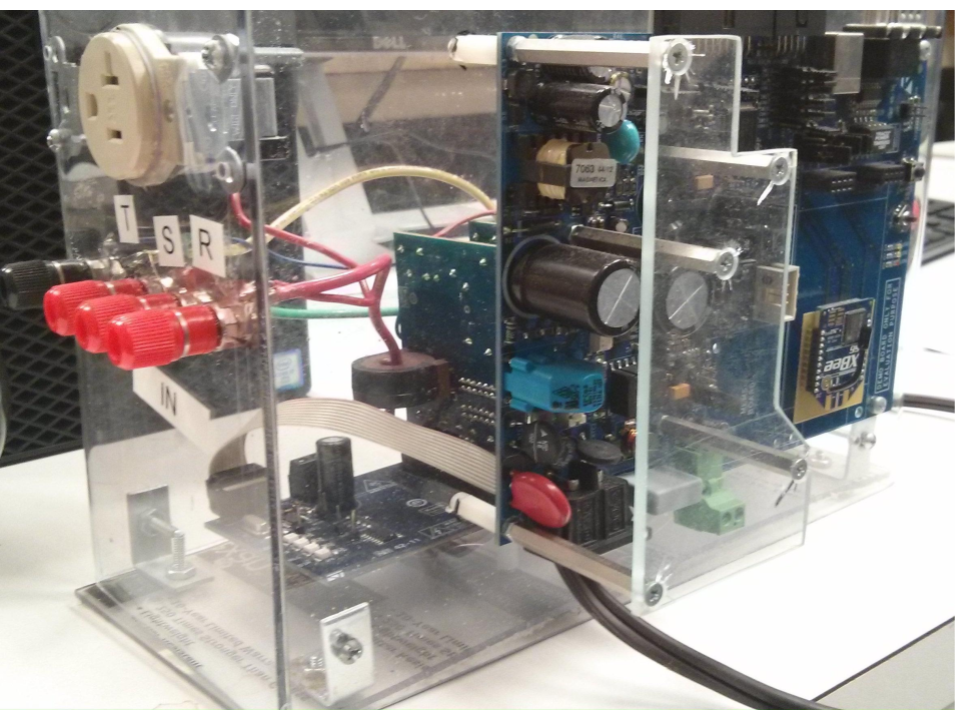}
\\[3mm]
\includegraphics[width=\columnwidth,height=1.6cm]{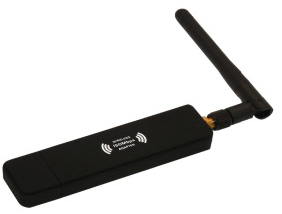}
\hspace*{10pt}\small{\textbf{(a)} Smart meter}
\end{minipage}
\begin{minipage}{0.7\columnwidth}
\includegraphics[width=\columnwidth,height=42mm]{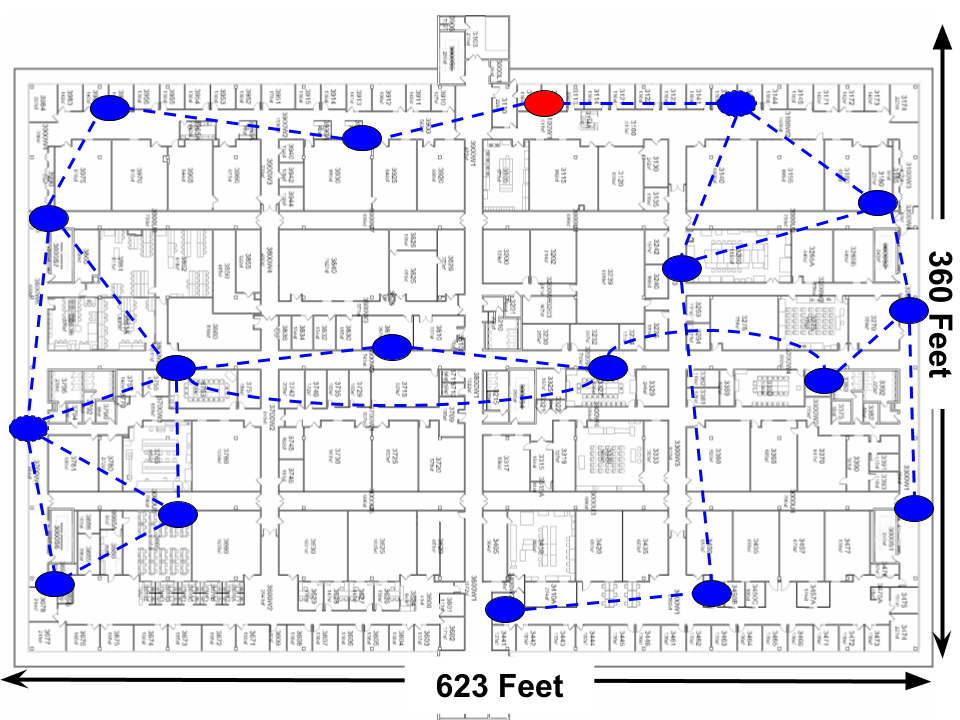}
\hspace*{40pt} \small{\textbf{(b)} Testbed topology}
\end{minipage}
\caption{AMI Testbed}
\vspace{-6pt}
\end{figure}

\begin{thisnote}
Although ns-3 provides very good simulation environment in terms of signal propagation, it still lacks to reflect the effects of real conditions on the signal such as path attenuation, refraction and diffraction while it propagates in wild. To see the effects of such conditions, we also built an IEEE 802.11s-based mesh network comprised of 18 Protronix  Wi-Fi  dongles attached  to  Raspberry-PIs which are integrated with the in-house meters  as shown in Fig. 4a. We carefully dispersed the meters on the floor as shown in Fig.~4b and build the shown multi-hop routing structure among meters by limiting transmission range by decreasing Tx-Power up to by a factor of 16 \citep{tourrilhes1997wireless}. By such positioning and decreased Tx-Power, we strive to mimic realistic conditions on signal propagation and its effects on multi-hop routing in a real AMI setup.
\end{thisnote}

\subsection{Communication Overhead}
\subsubsection{Distribution Overhead} 
\begin{wrapfigure}[9]{l}{0.65\textwidth}
\centering
  \vspace{-16pt}
   \centering
   \includegraphics[width=0.65\textwidth]{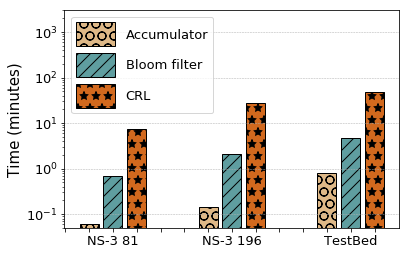}
    \vspace{-32pt}
    \caption{CRL distribution overhead}
    \label{fig:Completion_Time}
\end{wrapfigure}

In this subsection, we report on the completion time for the non-revoked proofs distribution of our approach with respect to other baselines both in simulation and testbed environments. The results which are shown in Fig.~\ref{fig:Completion_Time} indicate the accumulator approach significantly reduces the completion time compared to local CRL and bloom filter approaches due to condense accumulating. Even with respect to Bloom filter, which is touted as one of the most efficient methods in the literature, our approach reduced the completion time in approximately more than 10 orders of magnitude. 

Another critical observation from the simulation results is the scalability capabilities of our approach. While especially for the local CRL approach, the completion time increases significantly, this is not the case for our approach.
This can be attributed to the fact that the accumulator value is independent of the revoked CRL size while the overhead of other methods is proportional to the CRL size. The main overhead of our approach is directly related to the accumulator setting which was 2048 bits in our case. Therefore, even for very large-scale deployments that can have millions of meters, the overhead will not be impacted. In analyzing the experiments results for the testbed, we observe that the completion time takes more time even though the network size is much smaller. This is mainly because of the signal propagation issues such as path attenuation, refraction, interference from other devices, etc. within the building which does not exist in ns-3 simulations. Such issues cause more errors and packet loss and thus increase the re-transmissions to complete all packet distributions. In fact, the AMI infrastructure might have a similar challenge depending on the geographical location (e.g., urban vs rural environments) and thus the distribution of CRL will become even more critical. Therefore, our approach will be more suitable for such environments to reduce the impact from the wild.  
 
\subsubsection{Update Overhead}
\begin{wrapfigure}[9]{l}{0.66\textwidth}
\centering
  \vspace{-14pt}
   \centering
    \includegraphics[width=0.65\textwidth]{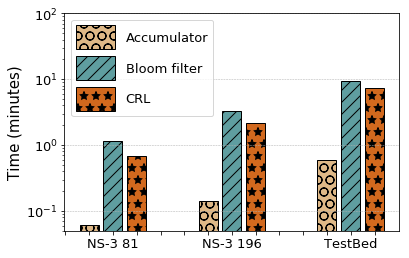}
  \vspace{-18pt}
  \caption{CRL update overhead}
  \label{fig:updateOverhead}
\end{wrapfigure}

In this subsection, we conducted experiments to assess the overhead
of CRL updates assuming that such updates are done regularly using the \textit{delta CRL} concept. Fig.~\ref{fig:updateOverhead} shows revocation update overhead in terms of  the completion time.
As in the case of full CRL, our approach significantly outperforms others due to of the size of the delta CRL. However, the results for the Bloom filter approach shows a different trend this time. It performs worse than the local CRL approach. This can be explained as follows:
For each updated revocation information, the Bloom filters must be created from scratch to carry both previous and newly revoked certificates. As a result, updating the CRL will take slightly more time than the whole CRL distribution for Bloom filter and thus will take more time than the local CRL approach. Note that the overhead of CRL distribution is proportional to the size of the delta CRL and thus the completion time follows a similar trend with the results in Fig. \ref{fig:Completion_Time}.

For the testbed results, we observe a similar which consistent with the simulations.  Again, the completion time is more due to signal propagation and interference issues. 
\subsection{Computation Overhead}\label{sec:benchmark}
We have demonstrated in the previous subsection that our approach significantly reduces the communication overhead. But, we need to also assess whether such a reduction introduces any major computational overhead. Thus, in this subsection, we investigated a detailed computational overhead of our approach. Specifically, we conducted two types of experiments: 1) We assessed the overhead of the computations due to the accumulation process in the Accumulator Manager. These experiments were conducted on a computer which has 64-bit 2.2GHz CPU with 10 hardware cores, and 32 GB of RAM assuming that these are reasonable assumptions for the computer that will act as the Accumulator Manager. Moreover, we also investigated whether some of these computations can be parallelized to reduce the computation times through multi-thread implementations further; and 2) We assessed the computation time for the $RevocationCheck()$ function in meters by implementing it in a Raspberry Pi (smart meter). 

\subsubsection{Overhead Results for the Accumulator Manager}\label{sec:accmulatormanager}
In this subsection, we present and discuss the overhead at the Accumulator Manager by considering the functions  below:

\noindent \textbf{Computing Prime Representatives}:
To assess the computational overhead of prime representative generation, we computed prime representatives for different set sizes. Note that since both the valid and revoked certificates' serial numbers are used in our approach, the input size can become huge when AMI scales. 
\begin{wrapfigure}[9]{l}{0.65\textwidth}
\vspace{-15pt}
\centering
\includegraphics[width=0.65\textwidth]{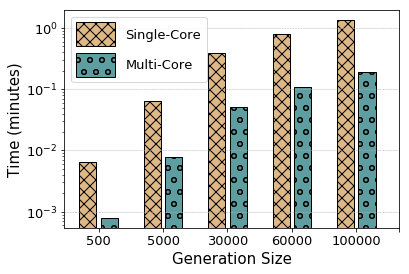}
\vspace{-35pt}
\caption{Prime representative computation}
\label{fig:primeOverhead}
\end{wrapfigure}
Therefore, we also conducted a benchmark test by using threads to show the parallelization ability of our approach. The results are shown in Fig.~\ref{fig:primeOverhead}.
As can be seen, the computational complexity of the prime representative generation is not overwhelming. $10^5$ representatives can be computed nearly in 1 minute even
using a single core. Parallelization reduces the computational complexity by roughly 10 folds which allows computational times in the order of seconds. 

\noindent \textbf{Computing the Accumulator Value}: 
Next, we benchmark the computation cost of accumulator value according to different CRL sizes as used in the previous experiment. In addition, we also conducted tests to assess the computational difference between our setting (i.e., the Accumulator Manager has all $aux_{info}$ information) and the case where the Accumulator Manager does not have $aux_{info}$ as discussed in Section IV.C. 
\begin{wrapfigure}[9]{l}{0.65\textwidth}
\vspace{-17pt}
\centering
\includegraphics[width=0.65\textwidth]{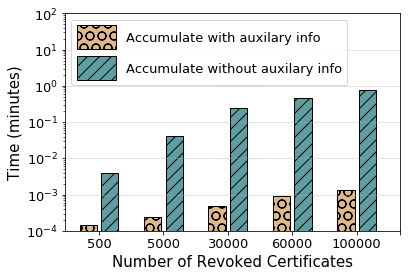}
\vspace{-36pt}
\caption{Accumulator computation}
\label{fig:accumulatorOverhead}
\end{wrapfigure} 
Note that for the computation of the accumulator value, a parallel implementation was not possible since each step in the computation depends on the previous operation. As seen in Fig.~\ref{fig:accumulatorOverhead}, the accumulator value is calculated under a minute for $10^5$ revoked certificates even without using $aux_{info}$. However, the availability of $aux_{info}$ significantly reduces the computation time making it possible to finish it milliseconds regardless of the size of the CRL.

\begin{wrapfigure}[9]{l}{0.65\textwidth}
\vspace{-16pt}
\centering
\includegraphics[width=0.65\textwidth]{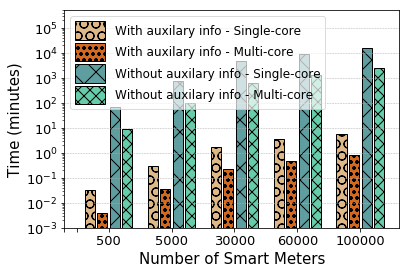}
\vspace{-38pt}
\caption{$nr_{proof}$ computation for \textit{full CRL}}
\label{fig:nonwitness1}
\end{wrapfigure}

\noindent \textbf{Computing Non-Revoked Values}: 
Finally, we assessed the overhead of the computation of non-revoked proofs for both the first setup phase by using \textit{full CRL} and the update phase by  using \textit{delta CRL}. Again, we conducted tests based on the availability/lack of  $aux_{info}$ and parallelization ability.Fig.~\ref{fig:nonwitness1} shows the computation overhead of this function according to different AMI sizes. As seen, $aux_{info}$ makes a significant difference in this case. 
Even with parallelization, the computational times are still in the order of days which may not be acceptable in an AMI setting. The results indicate that $aux_{info}$ needs to be available for efficient computations. We repeated the same experiment for the $UpdateNonRevokedProof()$ function and observed the same trends since the only change was the size of the CRL (i.e., delta CRL is much smaller). These results were not shown due to space constraints. 


\subsubsection{Overhead Results for Smart Meter}
\noindent \textbf{Revocation Check Overhead}:  We looked at the computational time of revocation check operations in smart meter based on the three approaches compared. This is an important experiment to understand the computation overhead of our approach on the smart meter, considering the fact that it has limited resources. 
As can be seen in Table~\ref{tbl:CRLCheck}, the elapsed time for a single revocation check is around 10 milliseconds in our approach. Comparing with the other methods, the Bloom Filter has the best results as expected  
because it enables faster checking by efficient hash operations. However, Bloom filter suffers from false-positives which degrades its efficiency by requiring access to the server \citep{akkaya2014efficient}. Our approach does not have such a problem. While our approach doubles the revocation check time compared to the local CRL method, the time is still pretty fast as it is in the order of milliseconds which does not impact any other operation. This is a negligible overhead given that it brings a considerable space-saving benefit which affects both distribution and storage overhead.
\begin{table}[h]
\setlength{\abovecaptionskip}{0cm}
\setlength{\belowcaptionskip}{0cm}
\renewcommand{\arraystretch}{1.1}
\caption{Elapsed Revocation Check Time}
\label{tbl:CRLCheck}
\centering
\begin{tabular}{ |m{3.0cm}|m{1.1cm}|m{1.1cm}|m{2.4cm}|  }
 \hline
 & \bf{Local CRL} & \bf{Bloom Filter}& \bf{Accumulator Approach}\\
 \hline 
 Average Time (ms) & 4.1 & 0.06 & 9.8 \\
 \hline 
  \end{tabular}
\end {table}

\noindent \textbf{Storage Overhead}: To compare the storage requirements, we identified the needed revocation information size for our approach and compared it with the other approaches, as shown in Table~\ref{tbl:storage}. As expected, accumulator has a superior advantage since smart meters just need to store a small accumulator value and non-revoked proof value. Local CRL, on the other hand, keeps  the whole CRL list and depending on the number of revoked certificates, it can be huge. For our scenario, the CRL size is around 0.7MB for 30K revoked certificates. While Bloom filter's performance is also promising, it is still not better than our approach and it suffers from false positives as discussed. 
\begin{table}[h]
\setlength{\abovecaptionskip}{0cm}
\setlength{\belowcaptionskip}{0cm}
\renewcommand{\arraystretch}{1.1}
\caption{CRL Storage Overhead}
\label{tbl:storage}
\centering
\begin{tabular}{ |m{3.5cm}|m{1.1cm}|m{1.1cm}|m{2.4cm}|  }
 \hline
 & \bf{Local CRL} & \bf{Bloom Filter}& \bf{Accumulator Approach}\\
 \hline 
 Required Space (MB) & 0.690 & 0.046 & 0.001 \\
 \hline 
  \end{tabular}
\end {table}

\section{Security Analysis}
\label{sec:security}
\begin{thisnote}
The security of the AMI depends on the secure implementation  of our approach. Therefore, we considered the threats in Section~\ref{sec:threat} against the security of the proposed approach and identified the relevant security goals. 

\begin{itemize}[label=\null]
\item \ballnumber{1} \textbf{Compromised smart meter attack:} 
An adversary can accomplish this attack in two different ways. For the first one, the adversary compromises the smart meter, and then the smart meter may be used to perform various attacks to Smart Grid. The UC is responsible for detecting malicious activity by utilizing different tools and sources. After detection, the UC puts the serial number of smart meter's certificate to the accumulation process. The updated $nr_{proof}$ and $\mathcal{A}$ values are distributed to the other smart meters, HES and gateways. This process basically detaches the compromised smart meter from the AMI. Every other non-compromised components which may interact with this smart meter are no longer be able to interact due to the our revocation check protocol. Once the information of the compromised smart meter is accumulated to obtain a new accumulator value, the attacker can not successfully bypass the revocation check mechanism by using the compromised smart meter itself or its stolen private key and certificate in the future. 

For the second one, the attacker can abuse a vulnerability in the certificate issuing process or steak smart meters' private keys from manufacturers. This time, the CA revokes certificates of the corresponding devices and publish new CRLs. Those CRLs collected by our CRL collectors. These newly revoked certificates are then accumulated, and the corresponding non-revocation proofs are disseminated to AMI. With the updated accumulator values, an adversary can not interact with any of the smart meters, HES, or gateways within AMI by using the revoked certificates.

\item \ballnumber{2} \textbf{Compromising the UC Servers:} In the event of an attack, the adversary's first target will be to compromise the accumulator manager to attack our revocation management. In our scheme, the accumulator manager plays a critical role but can be missed out easily because it is located within UC perimeters. However, the accumulator manager is the Achilles' heel of our approach and should be protected thoroughly. Thus, we investigate three possible attack scenarios and corresponding countermeasures within our revocation system.
\begin{enumerate}[label=\alph*)]
\item First, through the architectural design in Figure~\ref{fig:AMI_Inf} \emph{accumulator manager} is protected from any attacks by not allowing direct communication from outside of the network through two different firewalls. For instance, the second firewall configuration just allows incoming traffic, which is directly started by the Accumulator Manager to collect CRLs. Thus, it is not easy to access to accumulator manager from outside of the perimeter.
\item  Considering the ever-increasing threat environment and improved skills of adversaries, no matter what level of protection our system has, the accumulator manager can get compromised by breaking a path the proposed defense layers. In such a case, the key factor will be how quickly our approach responds to the incident. After detection of the compromise (e.g., attacker steals RSA setting parameters such as $aux\_info$ and $p\&q$), the accumulator manager can be migrated to another server, and new $nr_{proof}$ and $\mathcal{A}$  is computed from scratch by using different RSA primes $p'$ and $q'$ within minutes as shown in Section~\ref{sec:accmulatormanager}. Then, updated proofs are distributed to smart meters to prevent any further damage. So, our approach offers a pretty easy recovery capability which is important considering critical operations in AMI.
\item Third, our scheme is also allowing computation of $nr_{proof}$ without keeping critical security parameters of RSA accumulator settings  RSA (i.e., $aux_{info}$ and $p\&q$), since stolen $aux_{info}$ and $p\&q$ values enable a malicious actor to prove any arbitrary statements. These parameters can be deleted once they are used in the  setup phase. In such a case, compromising the accumulator manager does not give any advantage to the adversary to attack revocation management by abusing these parameters. Moreover, the computation of $nr_{proof}$ can still be accomplished for new smart meters or/and in case of updated revocation information, but it is more computationally intensive as shown in the Experiments Section.
\end{enumerate}

\item \ballnumber{3} \textbf{Compromising the Communication:}
As stated before, the traffic of AMI is generally unencrypted and causes additional attack surface to our approach. In this subsection, we investigate two possible attack scenarios and countermeasures against them as follows:
\begin{enumerate}[label=\alph*)]
\item \textbf{Accumulator freshness attack:}
An eavesdropping attack of AMI traffic poses a unique threat to our approach by combining public revocation information and circulating accumulator values. An attacker may perform a targeted attack if the UC has not updated the accumulator value properly by pinpointing the smart meters that use old accumulator values. However, our approach is robust to this attack since while computing the accumulator value, we use a secret prime number $r_k$ as a first exponent ($g^{r_k}$) in Eq.~\ref{eq:accumulator}. This prevents inferring the freshness of accumulator value by combining publicly known revocation information and circulated values in unencrypted traffic.

\item \textbf{Stolen non-witness attack:} One possible attack can be performed by using $nw_{1\&2}$ values to masquerade a valid smart meter since it can be obtained easily by eavesdropping. Our protocol is protected from this threat, even if the corresponding non-witness values $nw_{1\&2}$ of a smart meter are tried to be abused by attacker, the authentication during revocation check will still fail due to the multi-level signature checks. Thus, we relieve the attacks by abusing of stolen $nw_{1\&2}$ values through a multi-level authentication which combines the signature check with the accumulator check.
\end{enumerate}
\end{itemize}

\end{thisnote}

\section{Benefits and Limitations}
\begin{thisnote} 

There are several benefits associated with the use of the proposed approach for revocation management in AMI. Also, we highlighted some challenges and limitations of the approach.

\subsection{Benefits}
\begin{enumerate}
 \item \textit{Low overhead}: Our approach imposes minimal to no overhead to the smart meters deployed in the AMI and very low overhead to the central servers supporting the revocation management. In general, a revocation check contains at most one additional modular arithmetic operations if compared with the other revocation check methods (considering investigated methods realizes at least one modular arithmetic operation for signature check).  Also, the overhead imposed by disseminating the revocation information to the smart meters is very low. 
 \begin{table}[htb]
\label{tbl:applicability}
\centering
\caption{High-level Comparison of Revocation Management Schemes.}
\resizebox{0.95\textwidth}{!}{%
\begin{tabular}{lcccc}
\toprule
\multicolumn{1}{c}{{\color[HTML]{000000} }} &
  {\color[HTML]{000000} \textbf{Applicability}} &
  {\color[HTML]{000000} \textbf{\begin{tabular}[c]{@{}c@{}}Storage\\ Overhead\\Advantage\end{tabular}}} &
  {\color[HTML]{000000} \textbf{\begin{tabular}[c]{@{}c@{}}Communication\\
  Overhead\\Advantage\end{tabular}}} &
  {\color[HTML]{000000} \textbf{Security}} \\ \midrule
\rowcolor[HTML]{EFEFEF} 
{\color[HTML]{000000} OCSP}         & \Circle & \CIRCLE & \Circle & \CIRCLE \\
{\color[HTML]{000000} OCSP-Staple}  & \Circle & \CIRCLE & \Circle & \CIRCLE \\
\rowcolor[HTML]{EFEFEF} 
{\color[HTML]{000000} CRL}          & \LEFTcircle  & \Circle & \Circle & \CIRCLE \\
{\color[HTML]{000000} Delta CRL}    & \LEFTcircle & \CIRCLE & \LEFTcircle & \CIRCLE \\
\rowcolor[HTML]{EFEFEF} 
{\color[HTML]{000000} Bloom Filter} & \LEFTcircle & \CIRCLE & \LEFTcircle & \CIRCLE \\
{\color[HTML]{000000} Our Approach} & \CIRCLE & \CIRCLE & \CIRCLE & \CIRCLE \\
\bottomrule
\multicolumn{5}{c}{\small {\normalsize \CIRCLE} $=$ offers the benefit; {\normalsize \LEFTcircle} $=$ almost offers the benefit; {\normalsize \Circle} $=$ does not offer the benefit.} 
\end{tabular}%
}
\end{table}
 \item \textit{Applicability and Security}: The steps used in our approach can be easily implemented to the current AMI infrastructure with few adjustments. We showed that which components of the current AMI setup will be affected and need to be updated with new functionality.  To compare the applicability of our work with its alternatives, we determined four key benefits in total as shown in Table~3.  Our approach collects the revoked certificates information without interrupting the current smart grid operational network setup. However, unlike OCSP or OCSP-stapled methods, it requires extra communication overhead to distribute revocation information. Still, AMI communication infrastructure is not natural to an off-the-shelf OCSP-based solution due to the frequent query requirement, so obviously, it does not carry any advantage for decreasing the communication overhead.
On the other hand, our solution outperforms all other methods in terms of introduced distribution overhead. In brief, our conclusion from this comparative evaluation shows that our approach offers the same security benefits as other notable methods while keeping the overhead at the minimum level. 

\item \textit{A General Revocation Framework for Smart Grid}: Smart Grid is equipped with a myriad of various smart devices and sensors. This represents a new domain for security that is far beyond the traditional air-gapped operational network technology (OT) needs because of investments in distribution technologies such as renewable energy sources like rooftop solars and wind turbines. 
In this context,  our approach emerges as the first comprehensive solution that adapts the cryptographic accumulators to instrument lightweight revocation management and can be applied different domains in smart grid beyond AMI.
\end{enumerate}

\subsection{Limitations}
\begin{enumerate}
\item \textit{Tight Synchronization Requirement}: Cryptographic accumulators are powerful tools for short set representation and secure non-membership proofs. 
However, a disadvantage of using an accumulator-based revocation scheme is that the non-revoked proof and accumulator value has to be synchronized between smart meters. 
This might occur in two ways. First, the accumulator value at the verifier's site is out-of-date, but the non-revoked proof of the prover is updated and vice versa.
Asynchronous non-revoked proofs and accumulators between communicating smart meters may hinder the authentication operations; thus, AMI should ensure that all smart meters are updated and start to use the new proofs at the same time. Although the requirement for strict synchronization seems prohibitive, the AMI is a well-managed and synchronized network. Because of this characteristic of AMI, the synchronization requirement can be met easily.

\item \textit{Not-allow to use Unreliable Distribution Methods}: 
Another limitation related to synchronization requirement is that an attacker may selectively drop the packets to cause a synchronization problem between smart meters. Thus, any unreliable method for the distribution of non-revoked proofs should be avoided and the UC should ensure that required values are completely reached to smart meters.
\end{enumerate}

\end{thisnote}

\section{Conclusion}
Considering the overhead of certificate and CRL management in AMI networks, in this paper, we proposed a one-way cryptographic accumulator based approach for maintaining and distributing the revocation information. The framework condenses the CRLs into a short accumulator value and builds a \emph{secure}, efficient and lightweight revocation mechanism in terms of communication overhead. The approach is inspired by cryptographic accumulators and adopted based on the requirements of AMI.  
The experiment results indicate that the proposed approach can reduce the distribution completion time significantly for compared to CRL and Bloom filter approaches while introducing only minor additional computational overhead which is handled by the UC. There is no overhead imposed to smart meters.  
\begin{thisnote}
As future works, we first aim to incorporate an improved accumulator scheme to relax the tight synchronization requirement. Since different smart meters may use different proofs and accumulator values, the proposed method may require some architectural modifications to enable relaxed revocation check but still ensures security.
\end{thisnote}
Second, instead of using our in-house testbed, we will consider to use a real testbed such as EPIC \citep{adepu2018epic} to be able to test our approach on a realistic AMI infrastructure with integrated power grid components.
\section*{Acknowledgement}
This material is based upon work supported by  the Department of Energy under Award Number DE-OE0000779.
\vspace{2\baselineskip}

\section*{References}
\bibliography{mybib}
\end{document}